\documentclass[fleqn,a4paper,12pt]{article}
\usepackage{amssymb}
\usepackage{amsmath}
\usepackage[dvips]{graphicx}
\usepackage{placeins}
\usepackage{lscape}
\usepackage{a4}
\usepackage{epsfig}

\numberwithin{equation}{section}

\newcommand{\beq}{\begin{equation}} \newcommand{\eeq}{\end{equation}}
\newcommand{\bea}{\begin{array}} \newcommand{\eea}{\end{array}}
\newcommand{\ri}{{\mathrm i}}

\catcode`@=11 \long
\def\@caption#1[#2]#3{\par\addcontentsline{\csname
ext@#1\endcsname}{#1} {\protect\numberline{\csname
the#1\endcsname}{\ignorespaces #2}} \begingroup \small
\@parboxrestore \@makecaption{\csname fnum@#1\endcsname}
{\ignorespaces #3}\par \endgroup} \catcode`@=12

\renewcommand{\bar}{\overline}

\newcommand{\so}{\sigma_1}

\newcommand{\st}{\sigma_3}

\newcommand{\p}{\partial}

\newcommand{\R}{\mathbb{R}}

\DeclareMathOperator{\sech}{sech}
\DeclareMathOperator{\csch}{csch}

\catcode`@=11 \long
\def\@caption#1[#2]#3{\par\addcontentsline{\csname
ext@#1\endcsname}{#1} {\protect\numberline{\csname
the#1\endcsname}{\ignorespaces #2}} \begingroup \small
\@parboxrestore \@makecaption{\csname fnum@#1\endcsname}
{\ignorespaces #3}\par \endgroup} \catcode`@=12

\begin{document}
\allowdisplaybreaks
 \begin{titlepage}

 \vskip 2cm

\begin{center} {\Large\bf Generic matrix superpotentials} \footnote{E-mail:
{\tt nikitin@imath.kiev.ua},\ \ {\tt yuri.karadzhov@gmail.com}  }
\vskip 3cm {\bf {Anatoly G. Nikitin and Yuri Karadzhov} \vskip 5pt
{\sl Institute of Mathematics, National Academy of Sciences of
Ukraine,\\ 3 Tereshchenkivs'ka Street, Kyiv-4, Ukraine, 01601\\}}
\end{center}
\vskip .5cm \rm
\begin{abstract} A simple and  algorithmic description of matrix shape
invariant potentials is presented. The complete lists of generic
matrix superpotentials of dimension $2\times2$ and of special
superpotentials of dimension $3\times3$ are given explicitly. In
addition,  a constructive description of superpotentials realized by
matrices of arbitrary dimension is presented. In this way an
extended class of integrable systems of coupled Schr\"odinger
equation is classified. Examples of such systems are considered in
detail. New integrable multidimensional models which are reduced to
shape invariant systems via separation of variables are presented
also.

\end{abstract}
\end{titlepage}

\section{Introduction\label{intro}}
  Supersymmetry presents effective and elegant tools to solve quantum
  mechanical problems described by integrable Schr\"odinger equations.
  Unfortunately the class of known problems which can be solved using their
  supersymmetry is rather restricted since they should have the additional
  quality called shape invariance \cite{Gen}, and this feature
  appears to be rather rare. The classification of shape
  invariant (scalar)  potentials is believed to be completed at least
  in the case when they include an additive variable parameter \cite{Khare}.

  However there exist an important class of shape invariant
  potentials which is not classified yet, and they are matrix valued
  potentials. Such potentials appear naturally in models using
   systems of
  Schr\"odinger-Pauli equations. A famous example of such model was
  proposed by Pron'ko and Stroganov (PS)
\cite{Pron}, its supersymmetric aspects were discovered in papers
\cite{Vor} and \cite{Gol}. We note that there exist a relativistic
version of the PS problem which is shape invariant too \cite{ninni}.

Examples of matrix
  superpotentials including shape invariant ones were discussed in
  \cite{Andr}, \cite{Andri},  \cite{Ioffe} \cite{Rodr}, \cite{tkach}.
A rather general approach to matrix superpotentials was proposed in
paper \cite{Fu}, which, however,  was restricted to their linear
dependence on the variable parameter.

  A systematic study of
  matrix superpotentials was started  in recent paper \cite{yur1} where we presented
  the complete description of a special class of irreducible matrix
  superpotentials. These superpotentials include terms linear and inverse in
  variable parameter, moreover, the linear terms where supposed to be
  proportional to the unit matrix.  In this way we formulated five
  problems for systems of Schr\"odinger equations which are
  exactly solvable thanks to their shape invariance. Three of these
  problems are shape invariant with respect of shifts of two parameters,
  i.e.,  posses the dual shape invariance \cite{yur1}.

The present paper is a continuation and in some sense the completion
of the previous one. We classify all  irreducible matrix
superpotentials realized by matrices of dimension $2\times2$   with
linear and inverse dependence on variable parameter. As a result we
find 17 matrix potentials which are shape invariant and give rise to
exactly solvable problems described by Schr\"odinger-Pauli equation.
These potentials are defined up to sets of arbitrary parameters thus
the number of non-equivalent integrable models which are presented
here is rather large. They include as particular cases all
superpotentials discussed in \cite{Gol}, \cite {Ioffe}, \cite{Fu}
and \cite{yur1}, but also a number of new ones. Moreover, the list
of found shape invariant potentials is complete, i.e., it includes
all such potentials realized by $2\times2$ matrices.

In addition, we present a constructive description of
superpotentials realized by matrices of arbitrary dimension. The
case of matrix superpotentials of dimension $3\times3$ is considered
in more detail. A simple algorithm for construction of all
non-equivalent $3\times3$ matrix superpotentials is presented. A
certain subclass of such superpotentials is given explicitly.

The found superpotentials give rise to one dimensional integrable
models described by systems of coupled Schr\"odinger equations.
However, some of these systems are nothing but reduced versions of
multidimensional models, which appear as a result of the separation
of variables. Examples of such multidimensional systems are
presented in section 8. These systems are integrable and most of
them are new. In particular, we show that the superintegrable model
for vector particles, proposed in \cite{Pron2}, possesses
supersymmetry with shape invariance, and so its solutions can be
easily found using tools of SUSY quantum mechanics. The same is true
for the arbitrary spin models considered in paper \cite{Pron2}, but
we do not discuss them here.

We also analyze five of found integrable systems in detail and
calculate their spectrum and the related eigenfunctions. In
particular we give new examples of matrix oscillator models.

The paper is organized as follows. In section \ref{matrixproblem}
 we discuss restrictions imposed on superpotentials by
the shape invariance condition and present the determining equations
which should be solved to classify these superpotentials. In
sections \ref{2x2} and \ref{CompList} the $2\times 2$ matrix
superpotentials are described and the complete list of them  is
presented. The case of arbitrary dimensional matrices is considered
in section \ref{ArbDim}, the list of $3\times 3$ matrix
superpotentials can be found in section \ref{3x3}.

Sections \ref{models} and \ref{3d} are devoted to discussion of new
integrable systems of Schr\"odinger equations which has been
effectively classified in the previous sections. In addition, in
section \ref{3d} we discuss SUSY aspects of superintegrable models
for arbitrary spin $s$ proposed in \cite{Pron2}.

\section{Shape invariance condition\label{matrixproblem}}

Let us consider a Schr\"odinger-Pauli type equation
\begin{gather}\label{eq}H_\kappa\psi=E_\kappa\psi\end{gather}
where
\begin{equation}
\label{hamiltonian} H_k=-\frac{\partial^2}{\partial x^2}+V_k(x),
\end{equation}
and $V_k(x)$ is a matrix-valued potential depending on variable $x$
and parameter $k$.

We suppose that $V_k(x)$ is an $n\times n$ dimensional hermitian
matrix, and that Hamiltonian $H_k$ admits a factorization
\beq\label{s3}H_\kappa=a_\kappa^+a_\kappa^-+c_\kappa\eeq where
\[a_\kappa^-=\frac{\partial}{\partial x}+W_\kappa,\  \ a_\kappa^+=-
\frac{\partial}{\partial x}+W_\kappa,
\] $c_k$ is a constant and $W_k(x)$ is a
superpotential.

Let us search for superpotentials which generate shape invariant
potentials $V_k(x)$. It means that $W_k$ should satisfy the
following condition
\begin{gather}\label{SI}W_k^2+W'_k=W_{k+\alpha}^2-
W'_{k+\alpha}+C_k\end{gather}
 were $C_k$ and $\alpha$ are constants.

 In the following sections we classify shape invariant
 superpotentials, i.e., find matrices $W_k$ depending on
 of $x, k$ and satisfying conditions (\ref{SI}). More exactly, we find
 indecomposable hermitian matrices whose dependence on $k$ is
 is specified  by
 terms proportional to $k$ and $\frac1k$.

Let us consider  superpotentials of  the following generic form
\begin{equation}
\label{SP} W_k=k Q +\frac1k R+P
\end{equation}
where $P$, $R$ and $ Q$  are $n\times n$ Hermitian matrices
depending on $x$.

Superpotentials of generic form (\ref{SP}) were discussed in paper
\cite{yur1} were we considered the case of arbitrary dimension
matrices but restrict ourselves to the case when
 $Q=Q(x)$ is proportional to the unit
matrix. Rather surprisingly this supposition enables to make a
completed classification of  superpotentials (\ref{SP}). All such
(irreducible) superpotentials include known scalar potentials listed
in \cite{Khare} and five $2\times2$ matrix superpotentials found in
\cite{yur1}.

 To complete the classification presented in \cite{yur1} let us consider generic
 superpotentials (\ref{SP}) with arbitrary hermitian matrices
 $Q,\ P$ and $R$. We suppose  $W_k$ be irreducible, i.e., matrices
$R, P$ and $Q$ cannot be simultaneously transformed to a block
diagonal form.

 Substituting (\ref{SP}) into (\ref{SI}),
multiplying the obtained expression by $k^2(k+\alpha)^2$ and
equating the multipliers for same powers of $k$ we obtain the
following determining equations:
\begin{gather}
 Q'=\alpha( Q^2+\nu I),
\label{a0}\\\label{a00} P'-\frac\alpha2
\{ Q,P\}+\varkappa I=0,\\\label{a01} \{R,P\}+\lambda I=0\\
R^2=\omega^2I\label{a8}
\end{gather}
where $Q=\frac1\alpha \tilde Q,\ \ Q'=\frac{\p  Q}{\p x},\quad \{
Q,P\}= QP+P Q$ is an anticommutator of matrices $ Q$ and $P$, $I$ is
the unit matrix and $\varkappa, \ \lambda,\ \ \omega $ are
constants.

Equations (\ref{a0})--(\ref{a8}) have been deduced in \cite{yur1}
where the anticommutator $\{Q,P\}$ was reduced to doubled product of
$Q$ with $P$ since $Q$ was considered to be proportional to the unit
matrix and so be commuting with $P$.

The system (\ref{a0})--(\ref{a8}) for generic matrices $Q$, $P$ and
$R$ is much more complicated then in the case of diagonal $Q$.
However it is possible to find  its exact solutions for matrices of
arbitrary dimension.

\section{Determining equations for $2\times2$ matrix superpotentials\label{2x2}}
 At the first step we
restrict ourselves to the complete description of superpotentials
(\ref{SP}) which are matrices of dimension $2\times2$. In this case
it is convenient to represent $Q$ as a linear combination of Pauli
matrices
\begin{gather}\label{Qsig}Q=q_0\sigma_0+q_1\sigma_1+q_2\sigma_2+q_3\sigma_3\end{gather}
where $\sigma_0=I$ is the unit matrix,
 \begin{gather}\label{pm}
 \sigma_1=\begin{pmatrix}0&1\\1&0\end{pmatrix},\quad \sigma_2=\begin{pmatrix}0&-\ri\\\ri&0\end{pmatrix},\quad
 \sigma_3=\begin{pmatrix}1&0\\0&-1\end{pmatrix}.\end{gather}

 Let is show that up to the unitary transformation realized by a
constant matrix the matrix $Q$ can be transformed to a diagonal
form. Substituting (\ref{Qsig}) into (\ref{a0}) and equating
coefficients for the linearly independent Pauli matrices $\sigma_a,
a=1,2,3$ we obtain the following system:
\begin{gather}q_a'=2\alpha q_0q_a,\quad a=1,2,3,\label{q_a}
\end{gather}

It follows from (\ref{q_a}) that
\begin{gather}\label{q_a1}  q_a= c_aF(x), \quad F(x)=
\exp\left(2\alpha\int\!\!q_0 dx\right)\end{gather} where
  $c_a$ are integration
constants. Since all $q_a$ are expressed via the same functions of
$x$ multiplied by constants, we can transform $Q$ to the diagonal
form:
\begin{gather}\label{diag}Q\to
UQU^\dag=\left(\begin{array}{cc}q_+&0\\0&q_-\end{array}\right)\end{gather}
where $q_\pm=q_0\pm cF(x)$, $c=\sqrt{c_1^2+c_2^2+c_3^2}$, and $U$ is
the constant matrix:
\begin{gather*}U=\frac{c+c_3-\ri\sigma_2c_1+\ri\sigma_1c_2}{2\sqrt{c(c+c_3)}}
\end{gather*}

In accordance with (\ref{diag}) equation (\ref{a0}) is reduced to
the decoupled system of Riccati equations for $q_\pm$:
\begin{gather} \label{qq}q_\pm'=\alpha(q_\pm^2+\nu)\end{gather}
which is easily integrable. The corresponding matrices $P$ can be
found from equation (\ref{a00}):
\begin{gather}\label{PP}P=\left(\begin{array}{cc}p_+&p\\p^*&p_-\end{array}\right)\end{gather}
with $p_\pm$ being solutions of the following equation
\begin{gather}\label{PPd}p_\pm'=\alpha p_\pm q_\pm+\varkappa,\end{gather}
and
\begin{gather}\label{Pa}
p=\mu\exp\left(\frac12\alpha\int\!\!(q_++q_-) dx\right)\end{gather}
where $\mu$ is an integration constant and the asterisk denotes the
complex conjugation. Moreover, up to unitary transformations
realized by matrices commuting with $Q$ (\ref{diag}) the constant
$\mu$ can chosen be real and so we can restrict ourselves to $p$
satisfying $p^*=p$.

Consider the remaining equations (\ref{a01}) and (\ref{a8}).  In
accordance with (\ref{a8}) $R$ should be a constant matrix whose
eigenvalues are $\pm \omega$. Thus it can be represented as
$R=r_1\sigma_1+r_2\sigma_2+r_3\sigma_3$ where $r_1,\ r_2$ and $r_3$
are constants satisfying $r_1^2+r_2^2+r_3^2=\omega^2$, or
alternatively, $R=\pm\omega I$. Let $\omega\neq0$ then, in order
equation (\ref{a01}) to be satisfied we have to exclude the second
possibility and to set $r_1= \varkappa=p_\pm=0$. As a result we
obtain the general solution of the determining equations
(\ref{a0})--(\ref{a8}) with $\omega\neq0$ in the following form:
\begin{gather}\label{res}P=\sigma_1p,\quad
R=r_3\sigma_3+r_2\sigma_2,\quad
Q=q_+\sigma_++q_-\sigma_-\end{gather} where
$\sigma_\pm=(1\pm\sigma_3)/2,$ $q_\pm$ are solutions of Riccati
equation (\ref{qq}), $p$ is the function defined by (\ref{Pa}) and
$r_a$ are constants satisfying $r_2^2+r_3^2=\omega^2$.

If $\omega=0$ then conditions (\ref{a01}) and (\ref{a8}) became
trivial. The corresponding matrices $Q$ and $P$ are given by
equations (\ref{diag}) and (\ref{PP}).

\section{Complete list  of $2\times2$ matrix superpotentials\label{CompList}}
Let us write the found matrix superpotentials explicitly and find
the corresponding shape invariant potentials.

 All nonequivalent solutions of equations (\ref{qq})  are
enumerated in the following formulae:
\begin{gather}\label{lin1}\begin{split}&q_\sigma=0,\ \nu=0,\end{split}\\
\label{lin2}
\begin{split}&
q_\sigma=-\frac{1}{\alpha x+ c_\sigma}, \ \nu=0,\\&
q_\sigma=\frac\lambda\alpha\tan(\lambda x+ c_\sigma),\quad
\nu=\frac{\lambda^2}{\alpha^2}>0,\\&
q_\sigma=-\frac\lambda\alpha\tanh(\lambda x+ c_\sigma),\quad
\nu=-\frac{\lambda^2}{\alpha^2}<0,\\\end{split}\\\begin{split}&q_\sigma=-
\frac\lambda\alpha\coth(\lambda x+ c_\sigma),\quad
\nu=-\frac{\lambda^2}{\alpha^2}<0,\\&
 q_\sigma=-\frac\lambda\alpha,\quad
\nu=-\frac{\lambda^2}{\alpha^2}<0
\end{split}\label{lin3}
\end{gather}
where $\sigma=\pm, c_\sigma=\pm c$ and $c$ is an integration
constant.

Going over solutions (\ref{lin1})--(\ref{lin3}) corresponding to the
same values of parameter $\nu$ it is not difficult to find the
related entries of matrix $P$ (\ref{PP}) defined by equations
(\ref{PPd}) and (\ref{Pa}). As a result  we obtain the following
list of superpotentials:
\begin{gather}
\begin{split}&
W^{(1)}_\kappa=\lambda\left(\kappa\left(\sigma_+\tan(\lambda
x+c)+\sigma_-\tan(\lambda x-c)\right)\right.\\
&\left.+\mu\sigma_1\sqrt{\sec(\lambda x-c)\sec(\lambda
x+c)}+\frac{1}{\kappa}R\right),\end{split}\label{tan} \\
\begin{split}&
W^{(2)}_\kappa=\lambda\left(-\kappa(\sigma_+\coth(\lambda
x+c)+\sigma_-\coth(\lambda x-c))\right.\\
&\left.+\mu\sigma_1\sqrt{\csch(\lambda x-c)\csch(\lambda
x+c)}+\frac{1}{\kappa}R\right),\end{split}\label{cotanh1}
\\\begin{split}&
W^{(3)}_\kappa=\lambda\left(-\kappa(\sigma_+\tanh(\lambda
x+c)+\sigma_-\tanh(\lambda x-c))\right.\\
&\left.+\mu\sigma_1\sqrt{\sech(\lambda x-c)\sech(\lambda
x+c)}+\frac{1}{\kappa}R\right),\end{split}\label{tanh1}
\\\begin{split}&
W^{(4)}_\kappa=\lambda\left(-\kappa(\sigma_+\tanh(\lambda
x+c)+\sigma_+\coth(\lambda x-c))\right.\\
&\left.+\mu\sigma_1\sqrt{\sech(\lambda x+c)\csch(\lambda
x-c)}+\frac{1}{\kappa}R\right),\end{split}\label{tanhcotanh} \\
\begin{split}&
W^{(5)}_\kappa=\lambda\left(-\kappa(\sigma_+\tanh(\lambda
x)+\sigma_-)+\mu\sigma_1\sqrt{\sech(\lambda x)\exp(-\lambda
x)}+\frac{1}{\kappa}R\right),\end{split}\label{tanh_exp}
\\\begin{split}&
W^{(6)}_\kappa=\lambda\left(-\kappa(\sigma_+\coth(\lambda
x)+\sigma_-)+\mu\sigma_1\sqrt{\csch(\lambda x)\exp(-\lambda
x)}+\frac{1}{\kappa}R\right),\end{split}\label{cotanh_exp}
\\
\begin{split}&
W^{(7)}_\kappa=-\kappa
\left(\frac{\sigma_+}{x+c}+\frac{\sigma_-}{x-c}
\right)+\frac{\mu\sigma_1}{\sqrt{x^2-c^2}}+ \frac{1}{\kappa}R,
\end{split}\label{inin}
\\
\begin{split}&
W^{(8)}_\kappa=-\kappa \frac{\sigma_+}{x}
+\mu\sigma_1\frac{1}{\sqrt{x}}+ \frac{1}{\kappa}R
\end{split}\label{in0}\\\label{expp} W^{(9)}_\kappa= \lambda\left(-\kappa I+ \mu\exp(-\lambda
x)\so-\frac{\omega}{\kappa}\st\right).
\end{gather}
Here $\sigma_\pm=\frac12(\sigma_0\pm\sigma_3),\ \ R$ is the numeric
matrix given by equation (\ref{res}), $\kappa,\ \mu$ and $\lambda$
are arbitrary parameters.

 Formulae
(\ref{tan})--(\ref{in0}) give the complete list of superpotentials
corresponding to nontrivial matrices $R$. In particular this list
includes superpotentials with $Q$ being proportional to the unit
matrix which has been discussed in paper \cite{yur1}. These cases
are specified by equation (\ref{expp}) and equations (\ref{tan}),
(\ref{cotanh1}), (\ref{tanh1}), (\ref{inin}) with $c=0$ and
$R=\omega\sigma_1$.

Finally, let us add the list (\ref{tan})--(\ref{expp})  by
superpotentials with $R\equiv0$. Using equations (\ref{diag}),
(\ref{lin1})--(\ref{lin3}) and (\ref{PP}),  (\ref{Pa}) we obtain the
following expressions for operators (\ref{SP}):
\begin{gather}
\begin{split}&
W^{(10)}_\kappa=\lambda\left(\sigma_+(\kappa\tan(\lambda x+c)+\nu
\sec(\lambda
x+c))\right.\\
&\left.+ \sigma_-(\kappa\tan(\lambda x-c)+\tau \sec(\lambda x-c))+
\mu\sigma_1\sqrt{\sec(\lambda x-c)\sec(\lambda
x+c)}\right),\end{split}\label{tan0} \\ \begin{split}&
W^{(11)}_\kappa=-\lambda\left(\sigma_+(\kappa\coth(\lambda x+c)+\nu
\csch(\lambda
x+c))\right.\\
&\left.+ \sigma_-(\kappa\coth(\lambda x-c)+\tau \csch(\lambda x-c))+ \mu\sigma_1\sqrt{\csch(\lambda x-c)\csch(\lambda
x+c)}\right),\end{split}\label{cotanh10}
\\\begin{split}&
W^{(12)}_\kappa=-\lambda\left(\sigma_+(\kappa\tanh(\lambda x+c)+\nu
\sech(\lambda
x+c))\right.\\
&\left.+ \sigma_-(\kappa\coth(\lambda x-c)+\tau \csch(\lambda x-c))+ \mu\sigma_1\sqrt{\sech(\lambda x-c)\csch(\lambda
x+c)}\right),\end{split}\label{tanh10}
\\\begin{split}&
W^{(13)}_\kappa=-\lambda\left(\sigma_+(\kappa\tanh(\lambda x+c)+\nu
\sech(\lambda
x+c))\right.\\
&\left.+ \sigma_-(\kappa\tanh(\lambda x-c)+\tau \sech(\lambda x-c))+ \mu\sigma_1\sqrt{\sech(\lambda x-c)\sech(\lambda
x+c)}\right),\end{split}\label{tanhcotanh0} \\
\begin{split}&
W^{(14)}_\kappa=-\lambda\left(\sigma_+(\kappa\tanh\lambda x+\nu\sech
\lambda x) +\sigma_-\kappa)+\mu\sigma_1\sqrt{\sech\lambda
x\exp(-\lambda x)}\right),\end{split}\label{tanh_exp0}
\\\begin{split}&
W^{(15)}_\kappa=-\lambda\left(\kappa(\sigma_+(\kappa\coth\lambda
x+\nu\csch \lambda x +\sigma_-\kappa)+\mu\sigma_1\sqrt{\csch\lambda
x\exp(-\lambda x)}\right),\end{split}\label{cotanh_exp0}
\\
\begin{split}&
W^{(16)}_\kappa=-{\sigma_+}\left(
\frac{\kappa+\delta}{x+c}+\frac\omega2 (x+c)\right)-{\sigma_-}\left(
\frac{\kappa-\delta}{x-c}+\frac\omega2
(x-c)\right)+\frac{\mu\sigma_1}{\sqrt{x^2-c^2}},
\end{split}\label{inin0}
\\
\begin{split}&
W^{(17)}_\kappa=-{\sigma_+}\left(\frac{2\kappa+1}{2x}-\frac{\omega
x}{4} \right)+{\sigma_-}\left(\frac{\omega x}{2} +c\right)
-\mu\sigma_1\frac{1}{\sqrt{x}}.
\end{split}\label{in00}
\end{gather}

Formulae (\ref{tan})--(\ref{in00}) give the complete description of
matrix superpotentials realized by matrices of dimension $2\times2$.
These superpotentials are defined up to translations $x\rightarrow
x+c$,  $\kappa\rightarrow \kappa+\gamma$, and up to equivalence
transformations realized by unitary matrices. In
(\ref{tan})--(\ref{in00}) we introduce the rescalled  parameter
$\kappa=\frac{k}{\alpha}$ such that the transformations $k\to
k'=k+\alpha$ is reduced to:
\begin{gather}\kappa\to\kappa'=\kappa+1\label{kappa}.\end{gather}

The list (\ref{tan})--(\ref{in00}) includes all superpotentials
obtained earlier in \cite{Fu} and \cite{yur1}, but also a number of
new matrix superpotentials.  The corresponding shape invariant
potentials are easily calculated starting with superpotentials
(\ref{tan})--(\ref{in00}) and using the following definition:
\begin{gather}V_\kappa^{(i)}=W_\kappa^{(i)2}-W_\kappa^{(i)'},
\quad i=1,2,...,17.\label{ham}\end{gather}

To save a room we will not present all potentials (\ref{ham})
explicitly but restrict ourselves to discussions of particular
examples of them, see section \ref{models}.

\section{Matrix superpotentials of arbitrary
dimension\label{ArbDim}}

Let us consider generic
 superpotentials (\ref{SP}) with arbitrary dimension hermitian matrices
 $Q,\ P$ and $R$. In this case we again come to the determining equations
 (\ref{a0})--(\ref{a8}) where $Q, P$ and $R$ are now hermitian matrices of dimension $K\times K$ with arbitrary integer $K$.

 In accordance with (\ref{a8}) $R$ is a constant matrix whose eigenvalues are $\pm \omega$. Thus up to unitary equivalence it can be chosen in the form
\begin{gather}\label{R}R=\omega\left(\begin{array}{lc} I_n&0\\0&-I_m\end{array} \right),\quad n+m=K\end{gather}
where $I_n$ and $I_m$ are the unit matrices of dimension $n\times n$ and $m\times m$ correspondingly, $n$ and $m$ are the numbers of positive and negative eigenvalues of $R$.

It is convenient to represent  matrix $Q$  in a block form:
\begin{gather}\label{Q}Q=\left(\begin{array}{cc} A&B\\B^\dag&C\end{array}\right)\end{gather}
where $A, B$ and $C$ are matrices of dimension $n\times n$, $n\times
m$ and $m\times m$. Using the analogous representation for $P$ and
taking into account relations (\ref{a01}) we right it as
\begin{gather}\label{P}P=\left(\begin{array}{cc} 0&\hat P\\\hat P^\dag&0\end{array}\right)+\tau R\end{gather} where $\tau=-\frac{\lambda}{2\omega}$.

Substituting (\ref{R})--(\ref{P}) into (\ref{a0}) and (\ref{a00}) we obtain the following equations for the block matrices:
\begin{gather}\label{A} A'=\alpha(A^2+BB^\dag+\nu I_n),\\
C'=\alpha(C^2+B^\dag B+\nu I_m),\label{C}\\ \label{B}
B'=\alpha(AB+BC),\\\hat P'=\frac\alpha2(A\hat P+\hat PC),\label{Pe}\\
2\tau A+B\hat P^\dag+\hat PB^\dag=2\bar\mu I_n,\label{AB}\\-2\tau
C+B^\dag \hat P+\hat P^\dag B=2\bar\mu
 I_m\label{CB}\end{gather}
where $\bar\mu=\frac{\mu}\alpha$.

Thus the problem of description of matrix valued superpotentials
which generate shape invariant potentials is reduced to finding the
general solution of equations (\ref{A})--(\ref{CB}) for irreducible
sets of square matrices $A, C$ and rectangular matrices $B$ and $P$.
Moreover, $A$  and $C$ are hermitian matrices whose dimension is
$n\times n$ and $m\times m$ respectively  while dimension of $B$ and
$P$ is ($n\times m$). Without loss of generality we suppose that
$n\leq m$.

The system (\ref{A})--(\ref{CB}) is rather complicated, nevertheless
it can be solved explicitly. To save a room we shell not present its
cumbersome general solution but restrict ourselves to the special
subclass of solutions with trivial matrices $B$. In this case
$\bar\mu=\tau=0$ (otherwise the corresponding superpotentials are
reduced to a direct sum of $2\times2$ matrices considered in the
above), and the system is reduced to the following equations:
\begin{gather}\label{Aa} A'=\alpha(A^2+\nu I_n),\quad C'=\alpha(C^2+\nu
I_m),\\\hat P'=\frac\alpha2(A\hat P+\hat PC).\label{Pea}
\end{gather}

Without loss of generality the hermitian matrices $A$ and $C$ which
solve equations (\ref{Aa})  can be chosen as diagonal ones, see
Appendix. In other words their entries $A_{ab}$ and $C_{ab}$ can be
represented as:
\begin{gather}\label{A_C}A_{ab}=\delta_{ab}q_{b}, \quad C_{ab}=\delta_{ab}
q_{n+b}\end{gather}where $\delta_{ab}$ is the Kronecker symbol and
$q_\sigma$ ($\sigma=b $ or $\sigma=n+b$) are solutions of the scalar
Riccati equation
\begin{gather}q_\sigma'=\alpha(q_\sigma^2+\nu)\label{riki}\end{gather}
which is a direct consequence of (\ref{Aa}) and (\ref{A_C}).
Solutions of equations (\ref{riki}) are given by equations
(\ref{lin1})--(\ref{lin3}).

Thus matrices $A$, $C$ and $R$ are defined explicitly by relations
(\ref{Aa}), (\ref{lin1})--(\ref{lin3}) and (\ref{R})  while matrices
$B$ are trivial. The remaining components of superpotentials
(\ref{SP}) are matrices $P$ whose entries $\hat P_{ab}$ are easy
calculated integrating equations (\ref{Pe}):
\begin{gather}\label{PaB}\hat P_{ab}=\mu_{ab}\exp
\left(\frac12\alpha\int\!\!(q_a+q_{n+b}) dx\right)\end{gather} where
$\mu_{ab}$ are integration constants satisfying
$\mu_{ab}=(\mu_{ba})^*$, and $q_\sigma$ with $\sigma=a, n+b$  are
functions (\ref{lin1})--(\ref{lin3}) corresponding to the same value
of parameter $\nu/\alpha$.

In analogous way we can describe a special subclass of matrix
superpotentials (\ref{SP}) with trivial matrices $R$ \cite{Yura}. In
this case it is convenient to start with diagonalization of matrix
$Q$ and write its entries as
\begin{gather}Q_{\alpha\sigma}=\delta_{\alpha\sigma}q_\sigma,\quad
\alpha,\sigma=1,2,\dots,K\label{Q!}\end{gather} where $q_\sigma$ are
functions satisfying equation (\ref{riki}). Then the corresponding
entries of matrix $P$ satisfying (\ref{Pe}) are defined as:
\begin{gather}\label{PaBB} P_{\alpha\sigma}=\mu_{\alpha \sigma}\exp
\left(\frac12\alpha\int\!\!(q_\alpha+q_{\sigma})
dx\right).\end{gather} Functions $q_\alpha$ and $q_\sigma$ included
into (\ref{PaBB}) have to satisfy equation (\ref{riki}) with the
same value of parameter $\nu$. In addition, the matrix whose entries
are integration constants $\mu_{\alpha \sigma}$ should be hermitian.

\section{Superpotentials realized by $3\times3$ matrices\label{3x3}}

Let us search for superpotentials (\ref{SP}) realized by matrices of
dimension $3\times3$. We will restrict ourselves to the case when
parameter $\nu$ in the determining equation (\ref{a0}) is equal to
zero and find the complete list of the related superpotentials. Like
(\ref{inin}), (\ref{in0}), (\ref{inin0}) and (\ref{in00}) they are
liner combinations of power functions of $x+c_i$ with some constant
$c_i$.

There are three versions of the related matrices $R$ whose general form is given in (\ref{R}):
\begin{gather}R=\left(\begin{array}{cc}I_2&0\\0&1\end{array}\right),
\label{R1}\\R=I_3,\label{R2}\\
R=0_3.\label{R3}\end{gather}

Let us start with the case presented in (\ref{R1}). The
corresponding matrices $Q$ and $P$ are given by formulae (\ref{Q})
and (\ref{P}) with
\begin{gather}A=\left(\begin{array}{cc}a_1&a_2\\a^*_2&a_3\end{array}\right),
 \quad B=\begin{pmatrix}b_1\\
b_2\end{pmatrix}, \quad
\hat P=\begin{pmatrix}p_1\\
p_2\end{pmatrix}, \label{PU}\end{gather} where $a_1, a_2, a_3, c,
b_1, b_2, p_1$ and $p_2$ are unknown scalar functions. Moreover,
$a_1, a_3$ and $c$ should be real, otherwise $Q$ is not hermitian.
Without loss of generality we suppose that $p_1$ and $p_2$ be
imaginary, since applying a unitary transformation to $Q$, $P$ and
$R$ these functions always can be reduced to a purely imaginary
form. Moreover, the corresponding  transformation matrix is
diagonal.

In the previous section we {\it apriori} restrict ourselves to
trivial matrices $B$. Let us show that this restriction  is not
necessary at least for the considered case $\nu=0$.

First let us prove that system (\ref{A})--(\ref{CB}) is compatible
iff $\bar\mu=\tau=0$, and it is true for all versions of matrix $R$
enumerated in (\ref{R1})--(\ref{R3}). Calculating traces of matrices
present in (\ref{AB}) and (\ref{CB}) we obtain the following
relation:
\begin{gather}\label{Sp1}\tau
(\texttt{Tr}A+\texttt{Tr}C)+\bar\mu(m-n)=0\end{gather} where $m=2,
3, 0$ for versions (\ref{R1}), (\ref{R2}), (\ref{R1}) respectively,
and $n=3-m$.

 Differentiating all terms in
(\ref{Sp1}) w.r.t. $x$  and using equations (\ref{A}), (\ref{C}) we
obtain:
\begin{gather}\label{Sp2}\tau(\texttt{Tr}A^2+\texttt{Tr}C^2+2\texttt{Tr}B^\dag
B+\nu(m+n))=0.\end{gather}

Three the first terms in brackets are positive defined and we
supposed that $\nu=0$. If $\tau\neq0$ all terms in brackets should
be zero. If the trace of the square of hermitian matrix is zero,
this matrix is zero too, the same is true for matrix $B^\dag B$.
Thus for $\tau\neq0$  matrix $Q$ (\ref{Q}) is trivial. To obtain
non-trivial solutions we have to set $\tau=0$, then from (\ref{Sp1})
we obtain that $\bar\mu=0$ also.

 Substituting (\ref{PU}) into
(\ref{AB}) and (\ref{CB}) (remember that $\bar\mu=\tau=0$) we obtain
the following relations:
\begin{gather}\label{pb0}\begin{split}&p_1b^*_2-b_1p_2=0,
\quad p_2b^*_1-b_2p_1=0,\quad p_1(b_1-b^*_1)=0,\quad
p_2(b_2-b^*_2)=0.\end{split}\end{gather}

In accordance with (\ref{pb0}) there are three qualitatively
different possibilities:
\begin{gather}(a):\ p_1=p_2=0,\quad (b):\ p_1b_2=p_2b_1,\ b^*_1=b_1,\
b_2^*=b_2\label{a)}\end{gather} and
\begin{gather}(c):\ b_1=b_2=0.\label{c)}\end{gather}

In the cases (\ref{a)}) the corresponding superpotentials are
reducible. Indeed, in the case (a)  the only condition  we need to
satisfy is equation (\ref{a0}). But matrix $Q$ can be diagonalized
(see Appendix) and so the related superpotential can be reduced to
the direct sum of three scalar potentials.

The only possibilities to realize the case (b), which differs from
cases (a) and (c) is to suppose that
\begin{gather}p_1=\alpha p_2\quad
\text{and}\quad b_1=\alpha b_2\label{prop}\end{gather} or
$p_1=\alpha b_1$ and $p_2=\alpha b_2$ where $\alpha$ is a constant
parameter. The second possibility is excluded since $p_\alpha$ can
be proportional to $b_\alpha$ only in the case when these functions
are reduced to constants (otherwise $p_\alpha$ and $b_\alpha$ are
linearly independent, compare equations (\ref{B}) and (\ref{Pe})),
and so this possibility is reduced to (\ref{prop}) also.

But if conditions (\ref{prop}) are realized the corresponding
superpotential is reducible too since the transformation $W\to
UWU^\dag$ with
\[U=\frac1{\sqrt{1+\alpha^2}}\begin{pmatrix}\alpha&1&0\\-1&\alpha&0\\
0&0&\sqrt{1+\alpha^2}\end{pmatrix}\] makes it block diagonal.

Thus to obtain an irreducible superpotential (\ref{SP}) we should
impose the condition (\ref{c)}), and our problem is reduced to
solving the system (\ref{Aa}), (\ref{Pea}) where $\nu=0$. Like in
section \ref{2x2} the $2\times2$ matrix  $A$ can be chosen diagonal,
i.e., we can set $a_2=0$ in (\ref{PU}) while the remaining
(diagonal) entries of matrix $Q$ can be denoted as $a_1=q_1,
a_2=q_2, C=q_3$, compare with (\ref{A_C}). In accordance with
(\ref{riki}) with $\nu=0$,  functions $q_i$ can independently take
the following values:
\begin{gather}\begin{split}&q_1=-\frac1{x+c_1}\quad
\text{or}\quad q_1=0,\\& q_2=-\frac1{x+c_2}\quad \text{or}\quad
q_2=0,\\& q_3=-\frac1{x+c_3}\quad \text{or}\quad
q_3=0\end{split}\label{AC}\end{gather} where $c_1, c_2$ and $c_3$
are integration constants. The corresponding values of $p_1$ and
$p_2$ are easy calculated using equation (\ref{PaB}).  As a result
we obtain the following irreducible superpotentials:
\begin{gather}\begin{split}&W=(S_1^2-1)\frac\kappa{x+c_1}+(S_2^2-1)\frac\kappa{x+c_2}+
(S_3^2-1)\frac\kappa{x}\\&+ S_1\frac{\mu_1}{\sqrt{x(x+c_1)}}+
S_2\frac{\mu_2}{\sqrt{x(x+c_2)}}+\frac\omega\kappa (2S_3^2-1),\end{split}
\label{w1}\\
\begin{split}&W=(S_1^2-1)\frac\kappa{x}+(S_2^2-1)\frac\kappa{x+c}+
S_1\frac{\mu_1}{\sqrt{x}}+
S_2\frac{\mu_2}{\sqrt{x+c}}+\frac\omega\kappa
(2S_3^2-1),\end{split}\label{w2}
\\\begin{split}&W=(S_1^2-1)\frac\kappa{x+c}+(S_3^2-1)\frac\kappa{x}+
S_1\frac{\mu_1}{\sqrt{x}}+
S_3\frac{\mu_2}{\sqrt{x(x+c)}}+\frac\omega\kappa
(2S_3^2-1),\end{split}\label{w3}\\\begin{split}&W=(S_1^2-1)\frac\kappa{x}
+S_1c+ S_2\frac{\mu}{\sqrt{x}}+\frac\omega\kappa
(2S_3^2-1)\end{split}\label{w4}\end{gather} where $c, c_1, c_2, \mu,
\mu_1$ and $\mu_2$ are integration constants, and
\begin{gather}\label{s}S_1=\begin{pmatrix}0&0&0\\0&0&-\ri\\0&\ri&0\end{pmatrix}
,\quad
S_2=\begin{pmatrix}0&0&\ri\\0&0&0\\-\ri&0&0\end{pmatrix},\quad
S_3=\begin{pmatrix}0&-\ri&0\\\ri&0&0\\0&0&0\end{pmatrix}
\end{gather}
are matrices of spin $s=1$.

Formulae (\ref{w1})--(\ref{w4}) give the complete list of $3\times3$
matrix superpotentials including matrix $R$ in form (\ref{R1}). If
this matrix is proportional to the unit one (i.e., if the version
(\ref{R2}) is realized), the related matrix $P$ should be trivial,
see equation (\ref{a01}) with $\lambda=0$. Diagonalizing the
corresponding matrix $Q$ we obtain the direct sum of three scalar
potentials, i.e., the related superpotentials are reducible.

In the case (\ref{R3}) we again can restrict ourselves to diagonal
matrices $Q$ whose entries are enumerated in (\ref{AC}). The
corresponding  matrices $P$ can be calculated using equation
(\ref{PaBB}). As a result we obtain the following superpotentials:
\begin{gather}\begin{split}&W=(S_1^2-1)\frac\kappa{x+c_1}+(S_2^2-1)\frac\kappa{x+c_2}+
(S_3^2-1)\frac\kappa{x}\\&+ S_1\frac{\mu_1}{\sqrt{x(x+c_1)}}+
S_2\frac{\mu_2}{\sqrt{x(x+c_2)}}+
S_3\frac{\mu_3}{\sqrt{(x+c_1)(x+c_2)}},\end{split}\label{w5}\\
\begin{split}&W=(S_1^2-1)\frac\kappa{x}+(S_2^2-1)\frac\kappa{x+c}+
S_1\frac{\mu_1}{\sqrt{x}}+
S_2\frac{\mu_2}{\sqrt{x+c}}+S_3\frac{\mu_3}{\sqrt{x(x+c)}},\end{split}
\label{w6}\\\begin{split}&W=(S_1^2-1)\frac\kappa{x} +S_1c+
S_3\frac{\mu_1}{\sqrt{x}}+S_2\frac{\mu_2}{\sqrt{x}}.\end{split}\label{w7}\end{gather}

Formulae (\ref{w1})--(\ref{w7}) present the complete list of
irreducible  $3\times3$ matrix superpotentials corresponding to zero
value of parameter $\nu$ in (\ref{a0}).

In full analogy with the above we can find superpotentials with
$\nu$ nonzero. In this case the list of solutions (\ref{AC}) is
changed to solutions (\ref{lin1})--(\ref{lin3}) with $\sigma=1,2,3$
and the same value of $\nu$ for all values of $\sigma$. The
corresponding matrices $P$ again is calculated using equation
(\ref{PaB}) for non-trivial $R$ and equation (\ref{a01}) for $R$
trivial.

\section{Examples of integrable matrix models\label{models}}

Thus we obtain a collection of integrable models with matrix
potentials. The related superpotentials of dimension $2\times2$ are
given by equations (\ref{tan})--(\ref{in00}). They are defined up to
arbitrary constants $c, \lambda, \mu, \nu, \dots$. In addition, in
section \ref{ArbDim} we present the infinite number of
superpotentials realized by matrices of arbitrary dimension. So we
have a rather large database of shape invariant models, whose
potentials have the form indicated in (\ref{ham}).

Of course it is impossible to present a consistent analysis of all
found models in one paper. But we can discuss at least some of them.
In this and the following sections we consider some particular
examples of found models.

\subsection{Matrix Hamiltonians with Hydrogen atom spectra}
Let us start with the superpotential given by equation (\ref{inin}).
In addition to variable parameter $\kappa$ it includes an arbitrary
parameter $\mu$, and two additional parameters, $r_2$ and $r_3$,
which define matrix $R$ (\ref{res}). Moreover, $\mu^2+r_2^2\neq0$,
otherwise operator (\ref{inin}) reduces to a direct sum of two
scalar superpotentials.

The simplest version of the considered superpotential corresponds to
the case $\mu=0$ and $r_3=0, r_2=\omega$. Then with the unitary
transformation $W^{(8)}_\kappa\to W_\kappa= UW^{(8)}_\kappa U^\dag,
U=(1+i\sigma_3)/\sqrt{2}\ $ we transform $W^{(8)}_\kappa$ to the
following (real) form:
\begin{gather}W_\kappa=-\frac{\sigma_+\kappa}{x}-
\frac{\sigma_1\omega}{\kappa}. \label{simpl1}\end{gather}

The corresponding potential (\ref{ham}) looks as:
\begin{gather}V_\kappa=\frac{\kappa(\kappa-1)\sigma_+}{x^2}+
\frac{\omega\sigma_1}{x}+\frac{\omega^2}{\kappa^2}.\label{simpl2}\end{gather}

Shape invariance of potential (\ref{simpl2}) is almost evident.
Calculating its superpartner $V_\kappa^+=W_\kappa^2+W_\kappa'$ we
easily find that
\begin{gather}\label{shi} V_\kappa^+=V_{\kappa+1}+C_\kappa\end{gather}
where
\[C_\kappa=\frac{\omega^2}{(\kappa+1)^2}-\frac{\omega^2}{\kappa^2}.\]
Using this property (which makes our model be in some aspects
similar to the non-relativistic Hydrogen atom) we immediately find
spectrum of Hamiltonian (\ref{hamiltonian}) with potential
(\ref{simpl2}):
\begin{gather}E_N=-\frac{\omega^2}{N^2}\label{simpl3}\end{gather}
where $N=\kappa+n$, $n=0,1,2,\dots$

The ground state vector $\psi_{0}(\kappa,x)$ should solve the
equation
\begin{gather}a^-_\kappa\psi_{0}(\kappa,x)\equiv\left(\frac{\p}{\p
x}+W_\kappa\right)\psi_{0}(\kappa,x)=0\label{simpl4}\end{gather}
where $W_\kappa$ is $2\times2$ matrix superpotential (\ref{simpl1})
and $\psi_0(\kappa,x)$ is a two component function:
\begin{gather}\psi_0(\kappa,x)=\begin{pmatrix}\varphi\\
\xi\end{pmatrix}\label{psi00}.\end{gather} Substituting
(\ref{simpl1}) and (\ref{psi00}) into (\ref{simpl4}) we obtain the
following system:
\begin{gather}\varphi'-\frac{\kappa}{x}\varphi-\frac{\omega}{\kappa}\xi=0,
\label{simpl6}
\\\kappa\xi'-\omega \varphi=0.\label{simpl7} \end{gather}

Substituting $\varphi=\frac\kappa\omega\xi'$ obtained from
(\ref{simpl7}) into (\ref{simpl6}) we come to the second-order
equation for $\xi$:
\begin{gather*}\xi''-\frac\kappa{x}\xi'-
\frac{\omega^2}{\kappa^2}\xi=0,\end{gather*} whose normalizable
solutions are:
\begin{gather}\xi=\omega x^\nu K_\nu\left(\frac{\omega
x}\kappa\right),\quad \nu=\frac{\kappa+1}2\label{simpl8}\end{gather}
where $K_\nu(\cdot)$ are modified Bessel functions. The first
component of function (\ref{psi00}) is easy calculated using
(\ref{simpl7}):
\begin{gather}\label{simpl9}\
\varphi=\kappa(\kappa+1)x^{\nu-1}K_\nu\left(\frac{\omega
x}\kappa\right)-\omega x^\nu K_{\nu+1}\left(\frac{\omega
x}\kappa\right).\end{gather}

Solutions (\ref{psi00}), (\ref{simpl8}), (\ref{simpl9}) are square
integrable for any positive $\kappa$ and $0\leq x\leq\infty$.
Solution $\psi_n(\kappa,x)$ corresponding to the $n^\text{th}$
exited state can be obtained from the ground state vector using the
following standard relation of SUSY quantum mechanics (see, e.g.
\cite{Khare}):
\begin{gather}\label{psin}\psi_n(\kappa,x)=
a_{\kappa}^+a_{\kappa+1}^+ \cdots
a_{\kappa+n-1}^+\psi_0(\kappa+n,x).
\end{gather}

It is not difficult to show that vectors (\ref{psin}) are square
integrable too provided $\kappa$ is positive.

In analogous manner it is possible to handle ground state wave
functions corresponding to superpotential (\ref{inin}) with other
values of parameters $r_2, r_3$ and $\mu$. In general case (but for
$r_2\mu\neq0$) these wave functions are expressed via products of
exponentials $\exp\left(-\frac{\omega{x}}{\kappa}\right)$, powers of
$x$ and Kummer functions $U_{\alpha\nu} \left(\frac{2\omega
x}\kappa\right)$. We will not present the corresponding cumbersome
formulae here.

\subsection{Matrix Hamiltonians with oscillator spectra\label{oscil}}

Consider the next relatively simple model which corresponds to
superpotential (\ref{in00}) with $c=0$. Denoting
$W^{(17)}_\kappa=W_\kappa$ we obtain the following potential:
\begin{gather}\label{pot1}\begin{split}&V_\kappa=W_\kappa^2-W_\kappa'=
\sigma_+\left(\frac{4\kappa^2-1}{4x^2}+\frac{\omega^2x^2}{16}+
\frac{\mu^2}{x}-(\kappa+1)\frac{\omega}{2}\right)\\&+\sigma_-
\left(\frac{\omega^2x^2}{4}+\frac{\mu^2}{x}-\frac{\omega}{2}\right)-\sigma_1
\left(\frac{3\omega x}{4}
-\frac{2\kappa+1}{2x}\right)\frac{\mu}{\sqrt{x}}.\end{split}\end{gather}

It is easy to verify that the superpartner of $V_\kappa$, i.e.,
$V_\kappa^+=W_\kappa^2+W_\kappa'$ is equal to $V_{\kappa+1}$ up to a
constant term:
\begin{gather}\label{shape}V_\kappa^+=V_{\kappa+1}+\omega.\end{gather}
In other words, potential (\ref{pot1}) is shape invariant in
accordance with the definition given in \cite{Gen}. Moreover, like in the case of supersymmetric oscillator, the superpartner potential
$V_\kappa^+$ differs from $V_{\kappa+1}$ by the constant term $\omega$ which does not depend on variable parameter $k$.

Using standard tools of SUSY quantum mechanics it is possible to
find the spectrum of system (\ref{eq}), (\ref{hamiltonian}) with
hamiltonian (\ref{pot1}):
\begin{gather}E_n=n\omega,\quad n=0,1,2,...\label{osc}\end{gather}
which coincides with the spectrum of supersymmetric oscillator.

 The
ground state vector is defined as a square integrable solution of
equation (\ref{simpl4}). Substituting (\ref{in00}) and (\ref{psi00})
into (\ref{simpl4}) we obtain the following system:
\begin{gather}\varphi'-\left(\frac{2\kappa+1}{2x}-\frac{\omega
x}{4}\right)\varphi-\frac{\mu}{\sqrt{x}}\xi=0\label{1}\\\xi'+\frac{\omega
x}{2}\xi-\frac{\mu}{\sqrt{x}}\varphi=0.\label{2}\end{gather}

Changing in (\ref{1}), (\ref{2})
\begin{gather}\label{vp}\varphi=\exp\left(-\frac{\omega x^2}{4}\right)\tilde\varphi,\quad
\xi=\exp\left(-\frac{\omega x^2}{4}\right)\tilde\xi,\end{gather}
solving equation (\ref{2}) for $\tilde\xi$ and substituting the
found expression into (\ref{1}) we obtain the second order equation
for $\tilde\varphi$:
\begin{gather*}\tilde\varphi''-\left(\frac{\omega
x}{4}+\frac{\kappa}{x}\right)\tilde\varphi'-\frac{\mu^2}{x}\tilde\varphi=0.
\end{gather*}Its solutions are linear combinations of Heun
biconfluent functions:
$\tilde\varphi=c_1\tilde\varphi_1+c_2\tilde\varphi_2$ where
\begin{gather}\tilde\varphi_1=H_B(-a_+,0,a_-,b,cx),\quad
\tilde\varphi_2=x^{\kappa+1}H_B(a_+,0,a_-,b,cx)\label{3}\end{gather}where
\begin{gather}a_\pm=1\pm\kappa,\quad
b=\frac{4\sqrt{2}\mu^2}{\sqrt{\omega}},\quad
c=\frac{\sqrt{2\omega}}{4}. \label{4}\end{gather} Thus, in
accordance with (\ref{vp}),  (\ref{3}) and (\ref{1}) we have two
ground state solutions (\ref{psi00}) with
\begin{gather}\begin{split}&\label{5}\varphi=\varphi_1=\exp\left(-\frac{\omega x^2}{4}\right)
H_B(-a_+,0,a_-,b,cx),\\& \xi=\xi_1=\exp\left(-\frac{\omega
x^2}{4}\right)\left(\frac{\sqrt{2\omega
x}}{4\mu}H'_B(-a_+,0,a_-,b,cx)\right.\\&\left.-\frac1{2\mu}\left(\frac{2\kappa+1}{\sqrt{x}}+
\frac{\omega
x^\frac32}{2}\right)H_B(-a_+,0,a_-,b,cx)\right)\end{split}
\end{gather}
and
\begin{gather}\begin{split}&\label{6}\varphi=\varphi_2=x^{\kappa+1}H_B(a_+,0,a_-,b,cx),\\& \xi=\xi_2=\exp\left(-\frac{\omega
x^2}{4}\right)\left(\frac{\sqrt{2\omega
x}}{4\mu}H'_B(a_+,0,a_-,b,cx)\right.\\&\left.-\frac1{2\mu}\left(\frac{2\kappa+1}{\sqrt{x}}+
\frac{\omega
x^\frac32}{2}\right)H_B(a_+,0,a_-,b,cx)\right).\end{split}
\end{gather}

Functions (\ref{psi00}) whose components are defined in (\ref{5}) and
(\ref{6}) are square integrable for any real values of parameters
$\kappa, \mu$ and positive $\omega$.

Notice that for integer $\kappa$ solutions (\ref{5}) and (\ref{6})
are linearly dependent. We will not present here the cumbersome
expression of the second solution linearly independent with
(\ref{5}) which can be easy found solving system (\ref{1}),
(\ref{2}) for $\kappa$ integer.

One more matrix superpotential generating the spectrum of
supersymmetric oscillator is given by equation (\ref{inin0}).
Setting for simplicity $\delta=0$ we obtain the corresponding
Hamiltonian (\ref{ham}) in the following form:
\begin{gather}\begin{split}&V_\kappa=\frac{\omega^2}{4}(x^2+c^2)+
\frac{\mu^2}{x^2-c^2}+\left(\kappa+\frac12\right)\omega-
\sigma_1\mu x\left(\frac{2\kappa-1}{(x^2-c^2)^\frac32}+
\frac{\omega}{(x^2-c^2)^\frac12}\right)\\&+\kappa(\kappa-1)
\left(\sigma_+\frac1{(x+c)^2}+\sigma_-\frac1{(x-c)^2}\right)
.\end{split}\label{pot2}\end{gather}

Like (\ref{pot1}), potential (\ref{pot2}) satisfies the
form-invariance condition written in the form (\ref{shape}). The
spectrum of the corresponding Hamiltonian (\ref{hamiltonian}) is
given by equation (\ref{osc}). The ground state vectors, i.e.,
solutions of equation (\ref{simpl4}) where
$W_\kappa=W^{(16)}_\kappa$ is superpotential (\ref{inin0}) with
$\delta=0$, are given by equation (\ref{psi00}) with components
$\varphi$ and $\xi$ given below:
\begin{gather}\begin{split}&\varphi=\varphi_1=\exp\left(-4\omega
(x+c)^2\right)(c^2-x^2)^\kappa(c-x)^\frac12
H_C\left(a,b_-,b_+,d,r;\frac{x+c}{2c}\right),\\&\xi=\xi_1=\frac{\ri}{cx}
\exp\left(-4\omega (x+c)^2\right)(c^2-x^2)^\kappa\left(\frac{x-c}{2}
H'_C\left(a,b_-,b_+,d,r;\frac{x+c}{2c}\right)\right.\\&\left.+\left(\kappa+\frac12\right)
H_C\left(a,b_-,b_+,d,r;\frac{x+c}{2c}\right)\right)\end{split}\label{Sol2}\end{gather}
where $H_C(\dots)$ is the confluent Heun function,
\begin{gather}\label{Sol3}a=-4\omega c^2,\quad b_\pm=\kappa\pm\frac12,\quad
d=2\omega c^2,\quad
r=2b_+c^2\omega+\frac12\kappa^2+\frac38-\mu^2.\end{gather}

There exist one more ground state vector for Hamiltonian
(\ref{hamiltonian}) with potential (\ref{pot2}) whose components are
\begin{gather}\begin{split}&\varphi_2=\exp\left(-4\omega (x+c)^2\right)(c^2-
x^2)^\frac12(c-x)^\kappa
H_C\left(a,-b_-,b_+,d,r;\frac{x+c}{2c}\right),\\&\xi_2=\frac{\ri}{cx}
\exp\left(-4\omega
(x+c)^2\right)(c-x)^\kappa\left(c\left(2c\kappa-x\right)
H_C\left(a,-b_-,b_+,d,r;\frac{x+c}{2c}\right)\right.\\&\left.+\frac{x^2-c^2}{2}
H'_C\left(a,-b_-,b_+,d,r;\frac{x+c}{2c}\right)\right)\end{split}\label{Sol4}\end{gather}
where $a, b_\pm, d$ and $r$ are parameters defined by equation
(\ref{Sol3}).

For $\omega>0$ and $k>0$ functions (\ref{Sol2}) and (\ref{Sol4}) are
square integrable on whole real axis.
\subsection{Potentials including hyperbolic functions}
An important model of ordinary (scalar) SUSY quantum mechanics is
described by Schr\"o\-din\-ger equation with the hyperbolic Scarf
potential
\begin{gather}\label{Scarf}V_\kappa=-\kappa(\kappa-1)\sech^2(x).\end{gather}
This model possesses a peculiar nature at integer
values of the parameter $\kappa$, namely, it is a reflectionless
(non-periodic finite-gap) system which is isospectral with the free quantum mechanical particle. In addition, this model possesses
 a hidden (bosonized) nonlinear supersymmetry \cite{plush1}.

Let us consider shape invariant matrix potentials including
(\ref{Scarf}) as an entry. The corresponding superpotential can be
chosen in the form (\ref{tanh_exp}) where $\mu=0$ and $c_3=0,
c_2=\omega$. Then with the unitary transformation $W^{(7)}_\kappa\to
W_\kappa= UW^{(7)}_\kappa U^\dag, U=(1+i\sigma_3)/\sqrt{2}$ we
transform it to the real form:
\begin{gather}W_\kappa=\lambda\left(-\kappa(\sigma_+\tanh(\lambda
x)+\sigma_-)+\sigma_1\frac{\omega}{\kappa}\right).\label{tanh_exp1}\end{gather}
 The corresponding potential looks as:
\begin{gather}\label{pott1}V_\kappa=\lambda^2\left(-\sigma_+
\kappa(\kappa-1)\sech^2(x)-\sigma_1\omega(\tanh(\lambda x)+1)+\frac{\omega^2}{\kappa^2}+\kappa^2\right).
\end{gather}

Potential (\ref{pott1}) satisfies the shape
invariance condition (\ref{shi}) with
\begin{gather}\label{pott2}C_\kappa=\lambda^2\left(\frac{\omega^2}{(\kappa+1)^2}+(\kappa+1)^2-
\frac{\omega^2}{\kappa^2}-\kappa^2\right)\end{gather}
thus the discrete
spectrum of the  corresponding Hamiltonian (\ref{hamiltonian}) is
given by the following formula:
\[E=-\lambda^2\left(\frac{\omega^2}{(\kappa+n)^2}+(\kappa+n)^2\right)\]
where
\begin{gather}\label{pott3} n=0,1, \dots ,\quad  \kappa+n<0,\quad (\kappa+n)^2>\omega.\end{gather}
Conditions (\ref{pott3}) will be justified in what follows.

To find the ground state vector we should solve equation
(\ref{simpl4}) with $W_\kappa$ and $\psi_0(\kappa,x)$  given by formulae
(\ref{tanh_exp1}) and (\ref{psi00}) correspondingly. This equation is easy integrable and has the following normalizable solutions:
\begin{gather}\begin{split}&\varphi= y^{-\frac{\sqrt{\kappa^4+\omega^2}}
{\kappa}}(1-y)^{\frac\omega{2\kappa}-\frac\kappa2}{_2F_1}(a,b,c;
y),\\&\xi=\frac\kappa\omega\left(\kappa(2y-1)\varphi+2y(y-1)
\frac{\partial\varphi}{\p
y}\right).\end{split}\label{lab}\end{gather}
 Here
$_2F_1(a,b,c;y)$ is the hypergeometric function,
\begin{gather}\begin{split}&\label{tan11}c=1-\frac1\kappa\sqrt{\kappa^4+\omega^2},\quad
b=c+\frac\kappa2+\frac\omega{2\kappa}, \quad a=b-\kappa-1,\\& y=\frac12(\tanh\lambda x+1).\end{split}\end{gather}

Wave functions for exited states can be found starting with
(\ref{tan11}) and using equation (\ref{psin}). In order to functions
(\ref{tan11}) and the corresponding functions be square integrable,
parameters $\kappa, \omega$ and $n$ should satisfy condition
(\ref{pott3}), see discussion of normalizability of state vectors
including the hypergeometric function in section 10 of paper
\cite{yur1}.

Consider also superpotential (\ref{tanh_exp}) with $\omega=0$ and
$\mu\neq0$:
\begin{gather}\label{last0}W_\kappa=-\lambda\left(\kappa(\sigma_+(\tanh\lambda
x+\sigma_-)+\sigma_1\mu\sqrt{\sech\lambda x\exp(-\lambda x)}
\right).\end{gather}
 The corresponding potential (\ref{ham}) have the
following form:
\begin{gather}\begin{split}\label{last1}&V_\kappa=\lambda^2\left(-\sigma_+
\kappa(\kappa-1)\sech^2\lambda x
+\kappa^2\right.\\&\left.+\mu^2\sech\lambda x\exp(-\lambda x)
+\sigma_1\mu(2\kappa-1)\exp\frac{\lambda x}{2}\sech^{\frac32}\lambda
x\right).\end{split}\end{gather}

Solving equation
(\ref{simpl4}) with $W_\kappa$ and $\psi_0(\kappa,x)$  given by formulae
(\ref{tanh_exp1}) and (\ref{psi00}) we find components of the ground state vector:
\begin{gather*}\varphi=\frac{1}{\mu}\sqrt{\frac{1+\exp(2\lambda x)}{2}}
(\kappa\xi-\xi'),\quad\xi=y^\nu(1-y)^{-\frac\kappa2}\;{_2F_1}(a,b,c;
y),\\a=\nu-\frac\kappa2,\quad b=a+\kappa+\frac12,\quad c=1+2\nu,
\quad \nu=\frac12\sqrt{\kappa^2+2\mu^2},\quad y=\frac12(\tanh\lambda
x+1)\end{gather*} which are square integrable for $\kappa<0$. The
discrete spectrum   of Hamiltonian (\ref{hamiltonian}) with
potential (\ref{simpl2}) is given by the following formula:
$E=-\lambda^2(\kappa+n)^2$ where $n$ are natural numbers satisfying
the condition $\kappa+n<0$. If this condition is violated, the
related eigenvectors (\ref{psin}) are not normalizable.

\section{Multidimensional integrable models\label{3d}}

The models considered in previous subsections are one dimensional in
spatial variable. Of course it is more interesting to search for
multidimensional (especially, three dimensional) models which can be
reduced to integrable models by separation of variables. Famous
examples of such models are the (non-relativistic) Hydrogen atom and
the Pron'ko-Stroganov system \cite{Pron} which can be reduced to a
scalar and matrix shape invariant systems correspondingly. A more
"fresh" example is the  reduction of the AdS/CFT holographythe model
to the Poschl-Teller system proposed in \cite{plush2}.

In this section we consider new examples of the three-dimensional
Schr\"odinger-Pauli equations which can be reduced to a shape
invariant form by separation of variables. Moreover, the related
effective potentials in radial variable belong to the shape
invariant potentials deduced above.

\subsection{Spinor models \label{spinor}}

Consider shape invariant potential generated by the following
superpotential:
\begin{gather}W=(\mu\sigma_3-j-1)\frac1x+\frac\omega{2(j+1)}\sigma_1.
\label{3d1}\end{gather} This operator belongs to the list of matrix
superpotentials presented in section \ref{CompList}, see equation
(\ref{inin}). More exactly, to obtain (\ref{3d1}) it is necessary to
set $c=0, \kappa=j+1, r_2=0$ and $r_1=-\frac\omega2$ in
(\ref{inin}) and (\ref{res}). Then such specified superpotential
$W_\kappa^{(7)}$ appears to be unitary equivalent to $W$
(\ref{3d1}), namely,
\begin{gather}\label{trans}W=UW_\kappa^{(7)}U^{\dag}\quad
\text{with}\quad U=\frac1{\sqrt{2}}(1+\ri\sigma_2).\end{gather}

Calculating potential (\ref{ham}) corresponding to superpotential
(\ref{3d1}) with $\mu=\frac12$ we obtain:
\begin{gather}V=W^{2}-W'=\left(j(j+1)+\frac14-
\left(j+\frac12\right)\sigma_3\right)\frac1{x^2}
-\frac\omega{x}\sigma_1.\label{3d2}\end{gather} By construction,
potential (\ref{3d2}) is shape invariant, thus the related
eigenvalue problem
\begin{gather}\left(-\frac{\p^2}{\p
x^2}+V\right)\psi=E\psi\label{3d3}\end{gather} can be solved exactly
using standard tools of SUSY quantum mechanics. The corresponding
ground state vector is a two-component  function (\ref{psi00}) with
 \begin{gather}\label{3d4} \varphi=y^{j+\frac32}
K_{1}\left(y\right), \quad\tilde\xi=
y^{j+\frac32}K_{0}\left(y\right),\end{gather} where $y=\frac{\omega
x}{2(j+1)}$. Energy spectrum is given by equation (\ref{simpl3})
with $N=2j+n+1,\quad n=0,1,\dots$

The eigenvalue problem (\ref{3d3}) with potential (\ref{3d2})
includes the only independent variable $x$. However, it can be
treated  as a radial equation corresponding to the following three
dimensional Hamiltonian with the Pauli type potential:
\begin{gather}H=-\Delta+\omega{\mbox{{\boldmath $\sigma$}}}\cdot
{\bf B},\quad {\bf B}=\frac{{ \bf x}}{x^2}.\label{3d5}\end{gather}
Here $\delta$ is the Laplace operator, $\frac12\mbox{\boldmath{
$\sigma$}}$ is a spin vector whose components are Pauli matrices
(\ref{pm}), and ${\bf B}$ is the coordinate three vector divided by
$x^2=x_1^2+x_2^2+x_3^2$.

Of course, $\bf B$ has nothing to do with the magnetic field since
$\nabla\cdot{\bf B}\neq0$. However, it can represent another field,
e.g., the axion one \cite{wilczek}. Existence of such solutions for
equations of axion electrodynamics was indicated recently
\cite{Oksana}.

Expanding solutions of the eigenvalue problem for Hamiltonian
(\ref{3d5}) via spherical spinors we obtain exactly equation
(\ref{3d4}) for radial functions. We will not present the
corresponding calculation here which can be done using the standard
representations for the Laplace operator and matrix $\mbox{\boldmath
$\sigma$}\cdot \bf{x}$ in the spherical spinor basis, which can be
found, e.g., in \cite{FN}.

Let us remind that the  Pron'ko-Stroganov model is based on the
following (rescalled) Hamiltonian
\begin{gather}\label{ps}H=-\Delta+\frac{\sigma_1x_2-\sigma_2x_1}{r^2},\quad
r^2=x_1^2+x_2^2
\end{gather} which is reduced to the following form in cylindrical
variables:
\begin{gather}
\label{HamP} \hat{H}_m=-\frac{\partial^2}{\partial
r^2}+m(m-\sigma_3) \frac{1}{r^2}+\sigma_1\frac{1} {{r}}
\end{gather}
(we ignore derivatives w.r.t. $x_3$).

Hamiltonian (\ref{HamP}) can be expressed in the form (\ref{s3}).
Moreover, the corresponding superpotential again can be obtained
starting with superpotential (\ref{inin}) by setting
$\kappa=m+\frac12, \mu=\frac12, c=r_2=0, r_1=\frac12$ and making
transformation (\ref{trans}).
\subsection{Vector models\label{vector}}
Hamiltonian (\ref{ps}) corresponds to the physically realizable
system, i.e., the neutral fermion moving in the field of straight
line constant current. A natural desire to generalize this model for
particles with spin higher then $\frac12$ appears to be hardly
satisfied since  if we simple change Pauli matrices in (\ref{ps}) by
matrices, say, of spin one, the resultant model will not be
integrable \cite{Gol2}.

In paper \cite{Pron2} integrable generalizations of model
Hamiltonian (\ref{ps}) to the case of arbitrary spin have been
formulated. The price paid for this progress was the essential
complication of the Pauli interaction term present in (\ref{ps}).
However, there are rather strong physical arguments for such a
complication \cite{Pron2}, see also \cite{Beckers} for arguments
obtained in frames of the relativistic approach.

In this section we present a new formulation of the spin-one Pron'ko
model \cite{Pron2}. Doing this we perform the following goals: to apply our
abstract analysis of shape invariant matrix potentials to
a physically relevant system and to show that
 this model is shape invariant and so can be easily solved using SUSY
 technique.

Let us start with superpotential (\ref{w1}) with $c_1=c_2=\mu_2=0,\
\mu_1=1$. Making the unitary (rotation) transformation $W\to
UWU^\dag$ with $U=\exp\left(\ri S_2\pi/4\right)$ and changing the
notations $x\to r, \kappa\to m+\frac12$ we reduce it to the
following form:
\begin{gather}
\label{s4}
W=\frac{1}{r}S_3-\frac{\omega}{2m+1}\left(2S_1^2-1\right)-\frac{2m+1}{2x}.
\end{gather}
In addition, we transform the spin matrices to the Gelfand-Tsetlin form:
\begin{gather} S_1=\frac1{\sqrt2}\begin{pmatrix}0&1&0\\1&0&1\\0&1&0\end{pmatrix},\quad S_2= \frac\ri{\sqrt2}\begin{pmatrix}0&-1&0\\1&0&-1\\0&1&0\end{pmatrix},\quad S_3=\begin{pmatrix}1&0&0\\0&0&0\\0&0&-1\end{pmatrix}.\label{SPIN2}\end{gather}

The corresponding shape-invariant potential looks as
\begin{gather}
\label{HamPP} V_m=W^2+W'=\left((m-S_3)^2-\frac14\right)
\frac{1}{r^2}+\omega\left(2S_1^2-1\right)\frac{1}
{{r}}+\frac{\omega^2}{(2m+1)^2}.
\end{gather}

So far we simple represented one of the numerous supersymmetric toys
classified in the above. Now we are ready to formulate a two
dimensional model which generates the effective potential
(\ref{HamPP}). The corresponding Hamiltonian can be written as:
\begin{gather}H=-\frac{\p^2}{\p x_1^2}-\frac{\p^2}{\p
x_2^2}+\omega\frac{2({\bf S}\cdot{\bf H})^2-{\bf H}^2}{|\bf
H|}\label{IH}.\end{gather} Here $\mathbf H$ is the two-dimensional
vector of magnetic field generated by an infinite straight current;
its components are $H_1=q\frac{x_2}{r^2}$ and
$H_2=-q\frac{x_1}{r^2}$.

First we note that the last term in (\ref{IH}) is a particular case
of the interaction term found in \cite{Pron2}, see equations (15),
(21), (29) here for $s=1$ and $\beta_1=\beta_0$. However, we believe
that our formulation of this term is more transparent.

Secondly, introducing radial and angular variables such that
$x_1=r\cos\theta,\ x_2=r\sin\theta$, and expanding eigenfunctions of
Hamiltonian (\ref{IH}) via eigenfunctions $\psi_m$ of the symmetry
operator $J_3=\ri\left(x_2\frac{\p}{\p x_1}-x_1\frac{\p}{\p
x_2}\right)+S_3$ which can be written as: \beq\label{s1}
\psi_m=\frac{1}{\sqrt{r}}
\begin{pmatrix}\exp(\texttt{i}(m+1)\theta)\phi_1(r)\\
\exp(\texttt{i}m\theta)\phi_2(r)\\\exp(\texttt{i}(m-1)\theta)\phi_3(r)\end{pmatrix}\eeq
we come to the following hamiltonian in radial variables:
$H=-\frac{\p^2}{\p r^2}+V_m\label{IHH}$ where $V_m$ is the effective potential which coincides with (\ref{HamPP}). Thus  Hamiltonian (\ref{IH}) is shape invariant and its discrete spectrum and the corresponding eigenvectors are easily calculated using the standard tools of SUSY quantum mechanics.

To end this section we present one more integrable model for vector boson. This model is three dimensional in spatial variables and is characterized by the following Hamiltonian
\begin{gather}H=-\Delta +\omega\frac{2({\bf S}\cdot{\bf B})^2-{\bf B}^2}{|\bf
B|}\label{IHAH}.\end{gather} Here ${\bf B}$ is the three dimensional vector defined in (\ref{3d5}) and ${\bf S}$ is the matrix vector whose components are given in equation
 (\ref{SPIN2}).

Like (\ref{3d5}), Hamiltonian (\ref{IHAH}) corresponds to the shape
invariant potential in radial variables, which looks as
\begin{gather*}V=\left(j(j+1)+S_3^2-(2j+1)S_3\right)\frac1{x^2}+(2S_1^2-1)\frac{\omega}{x}.\end{gather*} The corresponding superpotential can be obtained from (\ref{s4}) changing $m\to j+\frac12$.

\section{Discussion}

In spite of that (scalar) shape invariant potentials had been
classified long time ago, there exist a great number of other such
potentials which where not known till now, and they belong to the
class of matrix potentials. The first attempt to classify these
potentials which we made in recent paper \cite{yur1} enabled to find
five types of them which are defined up to arbitrary parameters.
They give rise to new integrable systems of Schr\"odinger equations
which can be easily solved within the standard technique of SUSY
quantum mechanics.

In the present paper we present an infinite number of such
integrable systems. In particular we present the  list of
superpotentials realized by matrices of dimension $2\times2$, see
equations (\ref{tan})--(\ref{in00}).
 The main value of the list is its completeness, i.e., it
includes all superpotentials realized by $2\times2$ matrices which
correspond to  Schr\"odinger-Pauli systems (\ref{eq}) which are
shape invariant w.r.t. shifts of variable parameters.

In section \ref{ArbDim} an extended class of arbitrary dimension
matrix superpotentials is described. We do not present the proof
that this class includes all irreducible matrix superpotentials.
However such assumption looks rather
 plausible since by a consequent differentiation of equations (\ref{AB})
 and (\ref{CB}) with using conditions (\ref{A})--(\ref{Pe}) we can obtain
 an infinite number of algebraic compatibility conditions for system
 (\ref{A})--(\ref{CB}) which are nontrivial but can be satisfied asking for
 $B$ be equal to zero. An alternative solution of these compatibility
 conditions is $P=0$, but it leads to reducible superpotentials. Computing
 experiments with system (\ref{A})--(\ref{CB}) for the cases of $n\times n$
 matrix superpotential with $n\leq5$ also support the assumption $B=0$, see
 section \ref{3x3} where we did not make this assumption {\it a priori} but
 prove it. Notice that for $n=2$ this condition is not necessarily  satisfied,
 but it is seemed to be the only exceptional case.

  Nevertheless in section \ref{ArbDim} we consider the condition
$B=0$ as an additional requirement, which enables to find
superpotentials of arbitrary dimension in a straightforward way.

Thus we obtain an entire collection of integrable systems of
Schr\"odinger equations. Some examples of these models are
considered in section \ref{models} where we we find their energy
spectra and ground state solutions. Among them there are two
oscillator-like matrix models whose spectra are linear in the main
quantum number, see section \ref{oscil}.

One dimensional integrable  models classified in the present paper
are especially interesting in the cases when they can be used to
construct solutions of two- and three-dimensional systems. A perfect
example of such shape invariant system is the radial equation for
the Hydrogen atom. Thus an important task is to search
 for multidimensional (in particular, two- and three-dimensional)
 models which can be reduced to found
shape invariant systems  after separation of variables. Some results
of our search can be found in section \ref{3d} were we present new
integrable problems for two- and three-dimensional equations of
Schr\"odinger-Pauli type. In particular we discuss SUSY aspects of
the Pron'ko-Stroganov model generalized to the case of vector
particles (such generalization was proposed in paper \cite{Pron2}).
It happens that the spin-one model is shape invariant and so it can
be easily integrated using tools of SUSY quantum mechanics.

Except the case $s=1$ we did not discuss superintegrable models
proposed in \cite{Pron2} for arbitrary spin $s$. However, it is
possible to show that they are supersymmetric too.

Let us note that the results presented in section \ref{3d} can be
considered only as an advertisement, and we plane to present the
detailed discussion of integrable multidimensional  models in the
following publications.

\renewcommand{\theequation}{A\arabic{equation}} %
\setcounter{equation}{0}
\appendix
\section{Solutions of matrix Riccati equations}

The corner stone of our classification of matrix superpotentials is
the diagonalization of matrix Riccati equations (\ref{a0}) and
(\ref{Aa}). For completeness, we present here a simple and
constructive algorithm of such diagonalization for hermitian
matrices of any dimension while the case of $2\times2$ matrices was
already discussed in section \ref{2x2}.

Let us start with equation (\ref{a0}) and make the following change
 of the dependent variable $Q$:
\begin{gather}
\label{M}
 Q=M+q I
\end{gather}
where $I$ is the unit matrix and $q=q_\sigma$ is one of the
particular solutions (\ref{lin2}) of the scalar Riccati equation
(\ref{riki}) (the unnecessary subindex $\sigma$ can be omitted).
Thanks to the randomness of integration constants in solutions
(\ref{lin2})
 we always can suppose that the unknown matrix  $M$
be nondegenerated.

Substituting (\ref{M}) into (\ref{a0}) and using the identity $
(M^{-1})'=-M^{-1}M'M^{-1}$ we obtain the following {\it linear}
equation for $M^{-1}$:
\begin{gather}
\label{equN}
 (M^{-1})'=-\alpha (I+2q M^{-1}).
\end{gather}
Its solutions have the following generic form:
\begin{gather}
\label{solN}
 M^{-1}=\rho(x) I+\theta(x)C
\end{gather}
where $\rho(x)$ and $ \theta(x)$ are scalar real valued functions
whose exact expressions can be easily found for any particular
solution (\ref{lin2}), and $C$ is an arbitrary constant matrix.

In accordance with (\ref{solN}) matrix $M^{-1}$ is {\it
diagonalizable} (remember that  $M^{-1}$ should be hermitian). The
same is true for matrix $M$ and so matrix $Q$ (\ref{equN}) is
diagonalizable too.

In complete analogy with the above one can justify  the
diagonalizability of matrices $A$ and $C$ satisfying equations
(\ref{Aa}).

\end{document}